\documentstyle[11pt,paspconf]{article}

\begin{document}

\title{N-body simulations of interacting disc galaxies}
\author{E. Athanassoula}
\affil{Observatoire de Marseille, 2 Place le Verrier, 13248 Marseille
cedex 04, France}

\begin{abstract}
Disc galaxies can be substantially modified by close encounters and
mergers, since their discs are very responsive components. 
Close interactions can be held responsible for the formation of
bridges and tails, as well as for the formation of some bars,
asymmetries and grand design spirals. Bound clumps can form in the
tails, due to self-gravity, and could evolve to dwarf
galaxies. Off-centerings and asymmetries in the central parts of barred
galaxies can be made by off-centered and/or oblique impacts of
sufficiently massive and 
compact companions. Similar impacts, but preferably centered, on
non-barred galaxies can form 
ring galaxies. Companions on initially near-circular orbits can also
cause changes to the target disc as they spiral gradually inwards.
Low density companions are disrupted before reaching
the center of the target and their debris form a thick disc. On the
other hand most 
of the mass of the high density companions reaches the center, where
it may form a bulge, thus entailing evolution along the Hubble
sequence. Such companions thicken and expand the target disc and may
also destroy bars in it. If their initial orbital plane is at an angle
to the plane of the disc of the target, they can cause the latter to tilt
substantially, depending on their mass and initial inclination. 
\end{abstract}


\section{Introduction}

Galactic discs are very responsive to external forcing. Thus when a
disc galaxy interacts (or merges) with another galaxy a lot of
disc sub-structures can either be formed or destroyed. As examples let
me mention

$\bullet$ bridges and tails

$\bullet$ spiral structure

$\bullet$ bars

$\bullet$ off-centerings / asymmetries

$\bullet$ rings

$\bullet$ warps

$\bullet$ lenses

$\bullet$ bulges

$\bullet$ thick discs

In this review I will be selective, neglecting some items, and
developing only a few aspects of others. The choice of 
topic reflects to a large extent my own interests. 
In many cases I will focus on results obtained by the Marseille group
with our GRAPE systems. I will
furthermore confine myself to purely stellar models, neglecting gas,
star formation and their various consequences.

\section{Bridges and tails}

These are often seen in interacting pairs of disc galaxies and are
amongst the most spectacular results of 
interactions. Bridges link the two galaxies, while tails extend in the
opposite direction. Amongst the best known examples are the Antennae
(NGC 4038/39), the Mice (NGC 4676)  and the Atoms-for-Peace galaxy (NGC
7252), where the tails extend 
several tens of kpcs from the main bodies of the galaxies. A yet more
extreme case is the Super-antennae, where the tail extent is of the
order of 350 kpc (IRAS 19254-7245; Mirabel, Lutz \&
Maza 1991). 

Toomre \& Toomre (1972) convincingly demonstrated 
that gravity alone could account for the formation of such structures,
so that there was no need to invoke magnetic or other forces. The most
impressive examples are formed by direct passages, where the angular
velocity of the companion is, temporarily, equal to that
of some of the stars in the disc of the other galaxy. In this way
there is a broad ``near-resonance'', and the effect of the interaction
can be strong. 

Self-gravity can make further substructure in the tails. Bound clumps
can form, containing both stars and gas. Such clumps could evolve to
dwarf galaxies as discussed by Barnes \& Hernquist (1992) and, from
the observational point of view, e.g. by Schweizer (1978), by Hibbard
et al. (1994) and by Duc et al.
(1998). Contrary to the stars and gas, the 
halo material will not particularly concentrate into these clumps, so
that the ratio 
of halo to luminous material could be quite low in dwarfs created from
such clumps.

Since tails extend to large distances from the centers of their parent
galaxy they could in principle be used as probes for the dark matter
halo (Faber \& 
Gallagher 1979). Dubinski, Mihos \& Hernquist (1996) considered a
series of galaxy models with the same disc and bulge,
and halos of very different extents and masses. These models are
relatively similar in their inner parts but have very different values
of luminous-to-dark mass ratio. Simulations of interactions 
showed that the galaxies with less massive and less
extended halos form longer, more massive and more striking tails than
the galaxies with more massive and more extended halos. This can be
easily understood since the latter galaxies will result in
interactions with higher relative velocity, in which there will be
fewer disc particles in near-resonance with the angular velocity of
the tidal forcing. Furthermore the disc particles will need more
energy to climb out of the deeper potential well. Thus Dubinski, Mihos
\& Hernquist (1996) set an upper limit to the halo mass in their
specific model. Nevertheless one should not extrapolate this 
to set limits on the disc to halo mass ratio in general,
since this limit can depend strongly on the distribution of matter
within the two components (Barnes 1998, Springel \& White 1998).

\section{Spiral structure}

A close passage of a sufficiently massive companion can form a grand
design spiral in the disc of the target galaxy. The best known
example, and one of 
the most spectacular ones, is the spiral in M51, a galaxy interacting with
its close companion NGC~5195. Statistical studies (e.g. Kormendy \&
Norman 1979, Elmegreen \& Elmegreen 1982) 
have shown that M51 is not a unique case and that discs with grand
design spirals very often have close companions, as noted also by
Toomre \& 
Toomre (1972), who proposed that the origin of these spirals is tidal. A
large number of 
simulations have since shown that triggering by a close and sufficiently
massive companion can indeed lead to the formation of a two armed
trailing grand 
design spiral (e.g. Toomre 1981 and contribution to this meeting,
Hernquist 1990, Howard \& Byrd 1990, 
Sundelius 1991, Salo \& Laurikainen 1993),
while the physical mechanism responsible for it, swing amplification, has
been presented by Toomre (1981, and contribution to this meeting).

\section{Bars}

Although bars could be the result of an instability of isolated galactic
discs, tidal forces can trigger their formation (Noguchi 1987,
Guerin, Combes \& Athanassoula 1990, Noguchi 1996, Miwa \& Noguchi
1998). 

An interesting question in this context is whether triggered bars have
the same basic properties as bars developing in isolated disc
galaxies. The references mentioned above suggest that low-amplitude
tidal forcing can trigger the formation of bars whose properties 
are largely determined by the internal structure of the target
galaxy. The situation is more complicated for the case of high
amplitude forcing. For such cases the work by Miwa \& Noguchi (1998)
suggests that the 
properties of the driven bars are quite different from those of bars
growing spontaneously in isolated discs. More work is
necessary on this very interesting point.

Bars can not only be triggered by interactions, but can also be
destroyed by them (Pfenniger 1991; Athanassoula 1996b, hereafter
A96b). This will be
discussed in more detail in section \ref{sec:bar_destruct}

\section{Off-centerings / asymmetries}

Many galaxies are at some level asymmetric and some show strong
asymmetries, either in their outer 
regions, or in their inner parts, or in both. Typical examples are M101,
the LMC, NGC~4027 etc. In some cases the center of a
given component, e.g. the bar, does not coincide with that of the other
components (old disc, dynamical center, etc.). The formation of such
asymmetries in the outer parts can 
be a natural consequence of interactions. The formation of
off-centerings or asymmetries in the inner parts could either be due to
a mode, or be the direct result of an
interaction. For example a compact and sufficiently massive companion
hitting the inner parts of a barred disc galaxy can push the bar to
one side. Examples of such impacts can be seen in many of my
simulations and a few cases have been shown and briefly discussed by
Athanassoula, Puerari \& Bosma (1997, hereafter APB), Athanassoula
(1996a, hereafter A96a) and A96b. All these simulations are fully
self-consistent, i.e. not only the disc, but also the halo and the
companion are described by particles. The bar, once displaced, sloshes
around in the central part of the disc. If the galaxy is
centrally concentrated, dynamical friction will drive the bar  
very fast back to the central regions and also strip it of a substantial
part of its material. Using a spherical object rather than a bar,
Athanassoula, Makino \& Bosma (1997) tested how these processes depend
on the central concentration of the target. A bar is even more vulnerable
than a spherical object, since, if the passages occur at an
awkward angle with respect to bar major axis, it can lose its form. 

The fact that off-centered bars survive longer in less centrally
concentrated galaxies
than in more concentrated ones is in good agreement with the observation
that such asymmetries are mainly seen in late type systems (e.g. de
Vaucouleurs \& Freeman 1972, Odewahn
1996). Furthermore, when the bar is pushed
off-center in the simulations, a long one-armed spiral is formed (see
e.g. figure 3 of A96a), very reminiscent of structures observed in late type
off-centered barred galaxies such as NGC~4027 (de Vaucouleurs \& Freeman 1972).

\section{Ringed galaxies}

Three type of rings can be found in galaxies:

- Polar rings, which are nearly perpendicular to the disc of the
galaxy

- Rings in barred galaxies located at the main resonances

- Ringed galaxies

Here I will only briefly discuss the third type of rings. For the
other two types see e.g. the review by Athanassoula \& Bosma (1985).

Ringed galaxies can be formed from the central impact of a
sufficiently massive and compact companion on a target galaxy (Lynds
\& Toomre 1976; Theys \& Spiegel 1976; 1977, Toomre 1978). The
temporary extra gravitational attraction due to this companion causes
material in 
the target disc to move inwards. This is followed by a recoil. Thus
the material in the target disc starts large epicyclic oscillations,
whose period increases with distance from the center of the target. The
oscillations drift out of phase and orbits crowd together and
produce an expanding ring. This is a density wave propagating
outwards. Often the first ring is followed by a second one and in some
cases spokes form between the two. The best known example of a ring
galaxy showing all these features simultaneously is the Cartwheel
galaxy (A0035-324).

Several simulations have followed the first pioneering ones
(e.g. Huang \& Stuart 1988, Appleton \& James 1990) and the
results have been reviewed and discussed by Appleton \& Struck-Marcell (1996).
I will here only briefly present some results obtained by the
Marseille group, mainly by APB.
They verified that the rings are indeed density
waves, as had been predicted theoretically by Lynds \& Toomre (1976)
and Toomre (1978). They found that the expansion velocity of the ring
decreases steadily with time, both for the first and the second ring
and that the first ring expands faster than the second one. The
amplitude, the width, the lifetime and 
the expansion velocity of the first ring increase considerably with the
mass of the companion. The same is true for the velocity of the particles in
the ring. Rings formed 
by low mass encounters are more symmetric, more circular and
narrower. On the other hand encounters with high mass companions
increase substantially the extent of the target disc. As expected slow
impacts are more efficient and form first rings of higher relative amplitude,
with higher expansion velocities and longer lifetimes. 
The velocity of the particles in the ring is faster for slower
impacts. In general there is a broad correlation between the expansion
velocity of the first ring and the mean radial velocity of
the particles that constitute it at a given time. Perpendicular
impacts make more symmetric and circular rings than oblique ones,
which make more eccentric and broader rings. Finally such impacts make
substantial changes to the vertical structure of the disc.

\section{Can a companion destroy a bar without destroying the disc the
bar resides in?}
\label{sec:bar_destruct}

To answer this question I first tried simulations where the companion,
initially on a rectilinear orbit, was aimed
either perpendicularly, or at an angle at a barred disc
galaxy. In all my trials, however, either the companion was not
sufficiently massive, causing only a change in the bar pattern speed
and a drop in its amplitude, or,
when it was sufficiently massive, it destroyed the disc as well as the
bar (APB, A96a).  
Discouraged by these attempts, I tried a totally different type of
trajectory in which the companion started on a near-circular orbit, and I was
immediately rewarded by much more success (cf. also Pfenniger 1991).
 
The simulations that I will discuss here were all done on the GRAPE-3
and GRAPE-4 systems in Marseille Observatory. The former, together
with the available software and its performance, is discussed by
Athanassoula et al. (1998). For more information on the GRAPE systems
in general the reader is referred to Makino \& Taiji (1998).

I made two series of simulations. In the first series there are 120 000
particles in the target galaxy, while 
in the second series there are 800 000 particles in the
target, out of which 280 000 in the disc and the remaining in the
halo. A large fraction of the simulations in the first series were run
with a direct summation GRAPE code, while the remaining, as well as
the simulations in 
the second series, were run using the GRAPE treecode.
The first 
series consists of 46 simulations, and they include two different target
galaxies (with or without bulge), three different companions and seven
different orientations of 
the initial plane of the companion orbit. Only few simulations with
companions on retrograde orbits were tried. The second series consists
so far of twelve simulations, covering two different target discs,
three
different companions, and both direct and retrograde
companion orbits. In all cases of this second series, however, the
plane on the companion's 
orbit coincides with the plane of the target disc. The second series,
seen the number of particles used, was, to a large extent, aimed at a
study of the disc thickening and, more generally, to the changes of
the density and velocity distributions of the disc, the halo and the companion.
In both series the mass of all particles in a given
simulation is the same. Thus the ratio of masses of two components
is equal to the ratio of the number of particles in these two components. 
Initially the halo
was a Plummer sphere and the disc a Kuzmin/Toomre disc. I evolved the
disc galaxy in isolation until it developed a bar and chose as initial
condition for the interaction a time
when the bar was well developed. In this way the
companion will perturb an already barred galaxy. In other studies (e.g.
Pfenniger 1991), the companion is set in the simulation before the bar
has fully grown, so that the growth of the bar is partly stimulated by
the tidal force. This 
might cause problems with overshooting and thus complicates the
analysis. In order to measure the difference I started a number of
simulations in the first series with a non-barred disc, to allow
comparisons between spontaneous and stimulated bars.

The three companions considered have respectively a mass equal to 1.0,
0.29, and 0.1 times that of the disc.
All three have the same half mass radius and outer cut-off radius.
They start on near-circular orbits somewhat outside the outer edge
of the halo. Although it is quite consuming in CPU-time, such a start is
necessary, in order to avoid starting the simulation out of equilibrium (as it
would if the companion started within the halo) and to measure what
fraction of the companion's mass, energy and angular momentum stays in
the halo. 

Preliminary results from the first series of simulations are given by
A96b, while the tilt of the disc plane was discussed by both A96b and
Huang and Carlberg (1997). Here I will briefly present some further results,
mainly from the second series of simulations. 

\subsection{The fate of the companion: Bulge or thick disc}

The most massive companion is sufficiently dense not to be disrupted
by its passage through the halo and the disc. It loses only a small
percentage of its mass, while it spirals in towards the 
central regions of the target galaxy and most of it reaches the
center. Could it thus 
become the bulge of the target galaxy? Before asserting this I have
to examine the density and velocity distribution in the companion
after the merging and compare them with those of observed bulges. The
number of particles in the second series of simulations should be
sufficient for this task. If these detailed comparisons confirm first
impressions, then these simulations will argue that one possible way
of forming a bulge is by merging a target galaxy with a
sufficiently massive and compact companion. Thus such interactions, or
rather mergings, would entail evolution along the Hubble sequence,
since they would transform a late type galaxy with either a small
bulge, or no bulge at all, to an earlier type disc galaxy, with a sizeable
bulge component. 

Companions of the second (intermediate mass) and third type (smallest
mass) get disrupted well before reaching the center. They start
losing a substantial part of their particles when they reach the main
parts of the disc component. This is not done symmetrically
around the surface of the companion. Most of the particles leave from
the part of the companion that is farthest from the center of the
target, in a tail-like fashion. This ``tail'' winds more and more
tightly around the center of the target.
The structure is less clear with time, so that, after a sufficient time
has elapsed, the companion can be considered as forming a thick disk,
where some structure may or may not be visible. In all cases I
examined, this disc was quite thicker than the initial disc of the
target galaxy. This is presumably linked to the fact that the initial
diameter of the companion was in all simulations bigger than the
thickness of the target disc.

\subsection{The fate of the target disc}

The target disc suffers considerable changes during the interaction
and merger, which, as could be expected, can be seen clearest in the
case of the most massive companion. Indeed the disc suffers
considerable thickening, but also considerable extension in the radial
direction. The latter can be easily understood from the conservation of
angular momentum in the system, since initially the companion has
substantial angular momentum because of its high mass and its large
distance from the center of the target. As the target disc expands
both vertically and radially, its shape stays that of a disc. An
analysis of some of the simulations in the first series shows that the
disc suffers some small relative thickening, i.e. that the axial ratio
$c/a$ is larger after the merging than before. This needs to be
confirmed with the help of the second series of simulations. Indeed 
these measurements are not trivial, due to the presence of the bar,
which is thicker than the outer parts of the disc, due to the
formation of the peanut (Combes \& Sanders 1981, Combes et al. 1990,
Raha et al. 1991).

The most exiting development, however, is in the case where the orbit
of the companion is initially in a plane which does not coincide with
that of the target's disc, but is at an angle to it. Then the
target suffers a very dramatic tilt. In simulations with the highest mass
companion the plane of the target disc after the merging
is close to that of the initial orbital plane of the companion. This also
can be understood by angular momentum conservation. Such a plane
switch has not been considered in the analytical approach of Toth \&
Ostriker (1992). Since it absorbs a large fraction of the vertical
energy initially stored in the orbital motion of the companion, it
accounts for the fact that the target disc is not overly heated in the
vertical direction.

\subsection{The fate of the bar}

In the simulations where the companion was disrupted before reaching the
center of the target, the bar suffers some evolution but is not
destroyed. On the other hand in the simulations with the massive
compact companion, where most of the companion reaches the
target's center, the bar does not survive the merging. 
In the unperturbed barred system the particles sustaining 
the bar rotate around the center of the galaxy in orbits elongated
along the bar (i.e. orbits trapped around the $x_1$ family). When the
companion approaches the bar these orbits are severely perturbed, since
the mass of the companion is equal to that of the disc and therefore
larger than that of the bar. These important perturbations cause the
bar to be disrupted, as can be inferred by visualising the particles in
the disc at sufficiently frequent intervals during the last stages of
the interaction. Furthermore, after the merging is completed, the disc galaxy 
is much more centrally concentrated than it was at the beginning of
the simulation and this can stabilise it against bar instability
(A96b). The effect of central concentration on the bar instability has
been discussed at length for isolated galaxies e.g. by Hasan \& Norman
(1990), Hasan, Pfenniger \& Norman (1993), Friedli \& Benz (1993),
Friedli (1994) and Norman, Sellwood \& Hasan (1996).

\acknowledgments

I would like to thank Albert Bosma for
many useful discussions and Jean-Charles Lambert for his invaluable
help with the GRAPE software and the administration of the runs.
I would also like to thank IGRAP, the
INSU/CNRS and the University of Aix-Marseille I for funds to develop
the computing facilities used for the calculations in this paper.


\begin{references}
\reference Appleton, P. N., \& James, R. A. 1990, in Dynamics and
Interactions of Galaxies, 
ed. R. Wielen, Berlin: Springer Verlag, 200
\reference Appleton, P. N., \& Struck-Marcell, C., 1996, Fundamentals of
Cosmic Physics, 16, 111 
\reference Athanassoula, E. 1996a, in Barred Galaxies, 
R. Buta, B.G. Elmegreen 
\& D.A. Crocker, Astron. Soc. of the Pacific Conference Series,
309 (A96a)
\reference Athanassoula, E. 1996b, Barred Galaxies and Circumnuclear
Activity, Nobel Symposium No. 89, eds. 
    Aa. Sandqvist \& P.O. Lindblad, Lecture Notes in Physics, Vol. 474, 
    Springer Verlag, 59 (A96b)
\reference Athanassoula, E., \& Bosma, A. 1985, \araa, 23, 147
\reference Athanassoula, E., Bosma, A., Lambert, J. C., \& Makino, J. 1998,
\mnras, 293, 369
\reference Athanassoula, E., Makino, J., \& Bosma, A. 1997, \mnras, 286, 825
\reference Athanassoula, E., Puerari, I., \& Bosma, A. 1997, \mnras, 286,
284 (APB)
\reference Barnes, J. E. 1998, in Galaxy Interactions at Low and High
Redshift, J. E. Barnes \& D. B. Sanders, in press
\reference Barnes, J. E., \& Hernquist, L. 1992, Nature, 360, 715
\reference Combes, F., Debash, F., Friedli, D., \& Pfenniger, D. 1990,
\aap, 233, 82
\reference Combes, F., \& Sanders, R.H. 1981, \aap, 96, 164
\reference Dubinski, J., Mihos, J. C., \& Hernquist, L. 1996 \apj, 462, 576
\reference Duc P.-A., Mirabel, I.F. 1998, \aap, 333, 813
\reference Elmegreen, D. M., \& Elmegreen, B. G. 1982, \mnras, 201, 1021
\reference Faber, S. M., \& Gallagher, J. S. 1979, \araa, 17, 135
\reference Friedli, D. 1994, in Mass-Transfer Induced Activity in
Galaxies, ed. I. Shlosman, Cambridge: Cambridge Univ. press, 268
\reference Friedli, D., \& Benz, W. 1993, \aap, 268, 65
\reference Gerin, M., Combes, F., \& Athanassoula, E. 1990, \aap, 230, 37
\reference Hasan, H., \& Norman, C. 1990, \apj, 361, 69
\reference Hasan, H., Pfenniger, D. \& Norman, C. 1993, \apj,  
409, 91
\reference Hernquist, L. 1990, in Dynamics and Interactions of Galaxies,
R. Wielen, Berlin: Springer Verlag, 108
\reference Hibbard, J. E., Guhathakurta, P., Van Gorkom, J. H. \&
Schweizer, F. 1994, \aj, 107, 67
\reference Howard, S., \& Byrd, G. G. 1990 \aj, 99, 1798
\reference Huang, S., \& Carlberg, R. G. 1997, \apj, 480, 503
\reference Huang, S., \& Stewart, P. 1988, \aap, 197, 14
\reference Kormendy, J., \& Norman, C. A. 1979, \apj, 233, 539
\reference Lynds, R., \& Toomre, A. 1976, \apj, 209, 382
\reference Makino, J., \& Taiji, M. 1998, Scientific Simulations with
Special-purpose Computers, Chichester: Wiley pub. 
\reference Mirabel, I. F., Lutz, D., \& Maza, J. 1991, \aap, 243, 367
\reference Miwa, T., Noguchi, M. 1998, \apj, 499, 149
\reference Noguchi, M. 1987, \mnras, 228, 635
\reference Noguchi, M. 1996, \apj, 469, 605
\reference Norman, C., Sellwood, J. A., \& Hasan, H. 1996, \apj, 462, 114 
\reference Odewahn, S. C. 1996, Barred Galaxies, 
R. Buta, B.G. Elmegreen 
\& D.A. Crocker, Astron. Soc. of the Pacific Conference Series, 30
\reference Pfenniger, D. 1991, Dynamics of Disc Galaxies, B. 
Sundelius, G\"{o}teborg: G\"{o}teborg press, 191
\reference Raha, N., Sellwood, J. A., James, R. A., \& Kahn,
F. D. 1991, \aap, 352, 411
\reference Salo, H., \& Laurikainen, E. 1993, \apj, 410, 586
\reference Schweizer, F. 1978, in Structure \& Properties of Nearby
Galaxies, eds. E. Berkhuijsen \& R. Wielebinski, Dordrecht :
D. Reidel pub., 279 
\reference Springel, V., \& White, S. D. M. 1988, \mnras, in press (astro-ph/9807320)
\reference Sundelius, B. 1991, Dynamics of Disc Galaxies, B. 
Sundelius, G\"{o}teborg: G\"{o}teborg press, 195
\reference Toomre, A. 1978, in The Large Scale Structure of the Universe
M. S. Longair and J. Einasto, Dordrecht: Reidel pub., 109
\reference Toomre, A. 1981, in The Structure and Evolution of Normal
Galaxies, S. M. Fall \& D. Lynden-Bell, Cambridge: Cambridge Univ. press,
111
\reference Toomre, A., \& Toomre, J. 1972, \apj, 178, 623
\reference Toth, G., \& Ostriker, J. P. 1992, \apj, 389, 5
\reference Theys, J. C., \& Spiegel, E. A. 1976, \apj, 208, 650
\reference Theys, J. C., \& Spiegel, E. A. 1977, \apj, 212, 616
\reference de Vaucouleurs, G., \& Freeman, K. C. 1972, Vistas in
Astronomy, 14, 163 
\end{references}
\end{document}